\documentclass[aps,prx,groupedaddress, superscriptaddress,longbibliography,reprint,twocolumn]{revtex4-2}

\usepackage{amsmath}  % For math fonts, symbols, and environments
\usepackage{amssymb}
\usepackage{amsthm}
\usepackage{bbm}
\usepackage{graphicx}
\usepackage{color}
\usepackage{upgreek}
\usepackage[linkcolor = blue, citecolor = blue, urlcolor = blue, colorlinks = true]{hyperref}

\usepackage{amssymb}
\usepackage{bm}
\usepackage[capitalise]{cleveref}
\usepackage{textcomp}
\usepackage{hyperref}
\usepackage[usenames,dvipsnames]{xcolor}
\usepackage[normalem]{ulem}
\usepackage{wasysym}

\usepackage{floatrow}
\usepackage{wrapfig}
\usepackage{pgfgantt}

\usepackage{mathptmx}
\usepackage{bm}

\usepackage{fancyhdr}
\usepackage{blindtext}
\fancypagestyle{mypagestyle}{
\fancyhf{}
\fancyfoot[C]{\thepage}  %% change [C] to either [L] or [R] if needed.

}
\makeatletter
\let\ps@plain\ps@mypagestyle%% comment this to keep chapter page style
\makeatother
\begin{document}

%\preprint{APS/123-QED}

\title{Diffusiophoretic transport of colloids in porous media}% Force line breaks with \\

\author{Mobin Alipour}
\author{Yiran Li} 
\author{Haoyu Liu} 
\author{Amir A. Pahlavan}%
 \altaffiliation{Corresponding author}
 \email{amir.pahlavan@yale.edu}

\affiliation{%
 Department of Mechanical Engineering and Materials Science, Yale University, New Haven, Connecticut 06511, USA\\
 %This line break forced with \textbackslash\textbackslash
}%

\date{\today}% It is always \today, today,
             %  but any date may be explicitly specified

\begin{abstract}

Understanding how colloids move in crowded environments is key for gaining control over their transport in applications such as drug delivery, filtration, contaminant/microplastic remediation and agriculture. The classical models of colloid transport in porous media rely on geometric characteristics of the medium, and hydrodynamic/non-hydrodynamic equilibrium interactions to predict their behavior. However, chemical gradients are ubiquitous in these environments and can lead to the non-equilibrium diffusiophoretic migration of colloids. Here, combining microfluidic experiments, numerical simulations, and theoretical modeling we demonstrate that diffusiophoresis leads to significant macroscopic changes in the dispersion of colloids in porous media. We displace a suspension of colloids dispersed in a background salt solution with a higher/lower salinity solution and monitor the removal of the colloids from the medium.  While mixing weakens the solute gradients, leading to the diffusiophoretic velocities that are orders of magnitude weaker than the background fluid flow, we show that the cross-streamline migration of colloids changes their macroscopic transit time and dispersion through the medium by an order of magnitude compared to the control case with no salinity gradients. Our observations demonstrate that solute gradients modulate the influence of geometric disorder on the transport, pointing to the need for revisiting the classical models of colloid transport in porous media to obtain predictive models for technological, medical, and environmental applications.

\end{abstract}

%\keywords{Suggested keywords}%Use showkeys class option if keyword
                              %display desired
\maketitle

Chemical gradients drive the phoretic migration of colloids in a process known as diffusiophoresis \citep{Anderson89,Velegol16,Marbach19,Shin20b,Shim22,Ault24}. While sophisticated techniques for precise manipulation and guiding the self-assembly of colloids using external electric or magnetic fields have been developed over the years \citep{Whitesides02,Grzybowski09,Grzelczak10,Edwards14,Harraq22,Li22,Bishop23}, adding a ``pinch of salt" to control the motion of colloids and bio-macromolecules has proven to be an appealing alternative \citep{Abecassis08,Palacci10,Palacci12,Paustian13,Florea14,Kar14,Kar15,Shi16,Banerjee16,Shin16,Shin17,Shin17b,Shin17c,Nery17,Ault18,Lee18,Shin19,Ault19,Battat19,Banerjee19,Gupta19,Gupta20,Wilson20,Williams20,Shin20,Warren20,Shimokusu20,Singh20,Jotkar21,Shim21b,Alessio21,Alessio22,McKenzie22,Shim22b,Doan23,Lee23,Akdeniz23,Migacz23,Ghosh23,Lee23,Teng23,Yang23,Williams24,Migacz24,Yang24}. Beyond microfluidics, diffusiophoresis has been shown to influence the dispersion and clustering of the colloids in cellular, convective, chaotic and turbulent flows \citep{Volk14,Deseigne14,Schmidt16,Mauger16,Shukla17,Raynal18,Raynal19,Chu21,Volk22,Migacz22,Chu22,Anzivino24}, and has even been implicated in cargo transport, phase separation, and pattern formation within the crowded environment of the living cells \citep{Sear19,Ramm21, Burkart22, Alessio23,Shandilya24,Hafner24,Doan24,Shim24}.
 
Despite the ubiquity of chemical gradients in porous media flows \citep{Borch10,Li17,Rolle19,Dentz23}, from filtration and desalination membranes to microplastic spreading and drug delivery, the influence of diffusiophoresis on the dispersion and transport of colloids in porous media has remained mostly unexplored \citep{Shin18,Park21,Doan21,Tan21,Sambamoorthy23,Somasundar23,Jotkar24}. The classical framework describing colloid transport in porous media either treats colloids as infinitesimal tracers, whose trajectory is determined by the flow velocity and diffusion, or accounts for the finite colloid size and hydrodynamic or non-hydrodynamic interactions with the surrounding surfaces \citep{Saffman59,Brenner80,Koch85,Koch89,Elimelech90,Edwards91,Liu95,Ryan96,Johnson96,Roy97,Souto97,Kretzschmar99,Bradford02,Hoek03,Tufenkji04,Jonge04,Auset04,Kanti06,Bradford08,Molnar15,Miele19,Molnar19,Bizmark20,Fan22,Li22b,Patino23,Wu23,Wu24,Storm24}. In this framework, the influence of solutes is merely in determining the range of equilibrium, and short-ranged, DLVO-type interactions between the colloids and surfaces, and the macroscopic dispersion of colloids is described as a function of the mean flow velocity in the form of Peclet number \citep{Dentz18,Aramideh19,Puyguiraud21,Mangal21,Mangal21b,Mangal22,Kumar22,Darko24}. The classical description for colloid transport in porous media does not take into account the non-equilibrium diffusiophoretic effects.

\begin{figure*}%[tbhp]
\centering
\includegraphics[width= 1 \textwidth]{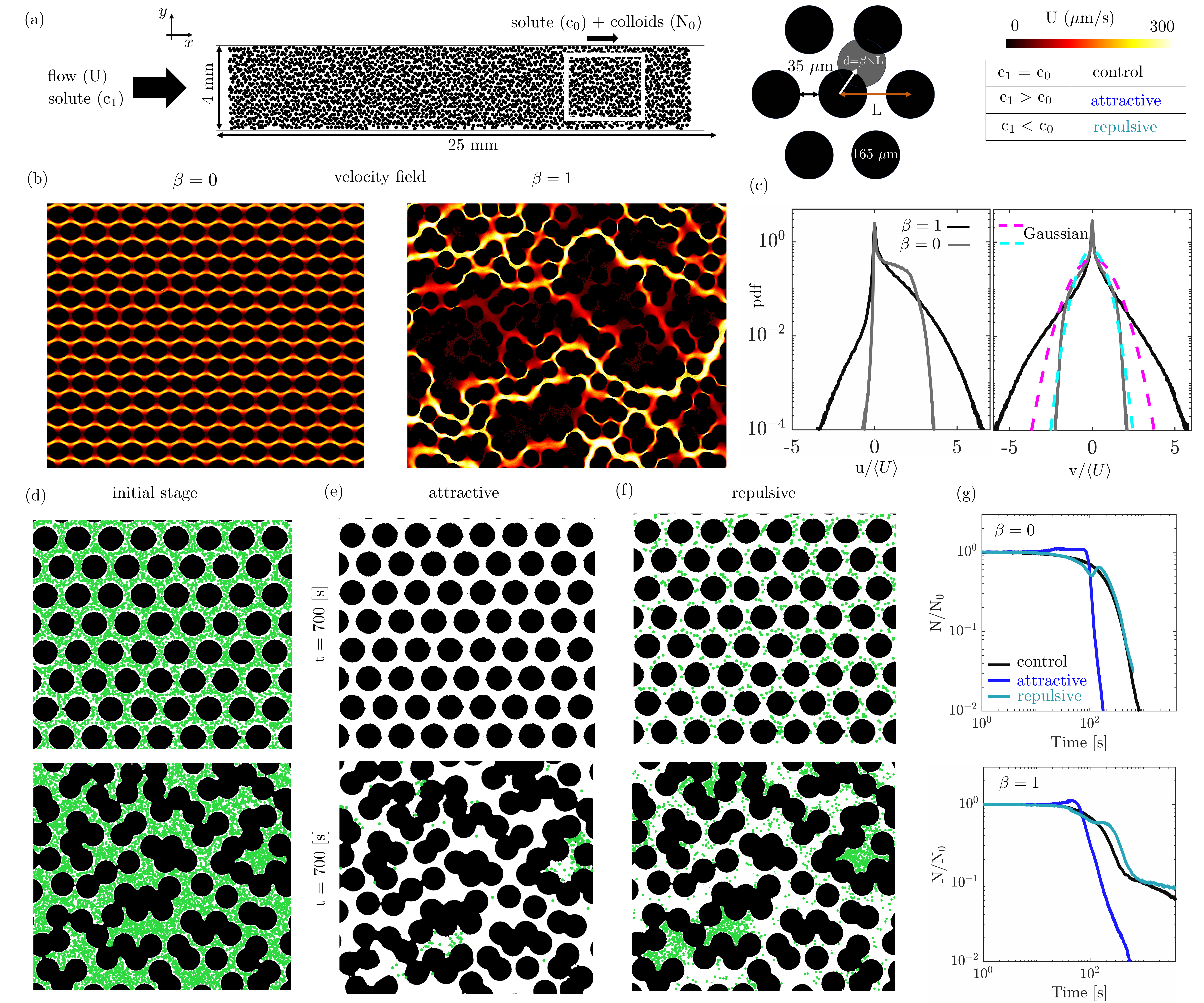}
\caption{Flow disorder and solute gradients shape the macroscopic transport of colloids in porous media. (a) Schematic of the microfluidic chips patterned with obstacles. The field of view (FOV) of the experiments is marked with the white rectangle. We introduce geometric disorder into an ordered lattice by randomly displacing the obstacles with an amplitude $\beta L$, where $L$ represents the center to center spacing between the obstacles in the ordered lattice. (b) While flow in the ordered lattice ($\beta=0$) is periodic, geometric disorder ($\beta=1$) leads to the emergence of preferential flow pathways surrounding low-velocity/stagnant fluid pockets. (c) The geometric disorder therefore broadens the probability distribution of both the longitudinal and transverse components of the velocity field, and leads to the deviation of the transverse component from the Gaussian distribution. (d-f) Displacing the initially uniformly dispersed colloids in both ordered and disordered media with a front with higher (attractive)/lower (repulsive) solute concentration, we then monitor how colloids get removed from the medium. (g) The time evolution of normalized number of colloids, $N(t)/N_0$, clearly demonstrates the influence of solute gradients on the macroscopic transport of colloids. Here, we have $c_1$/$c_0\approx 100$ in the attractive case, and $c_1$/$c_0\approx 0.01$ in the repulsive case. }
\label{fig1}
\end{figure*}

Here, we address the question of how solute gradients and flow disorder conspire to modulate the transport and dispersion of colloids in porous media. We fabricate microfluidic chips patterned with an ordered array of obstacles and introduce disorder by randomly perturbing their position. The amplitude of the perturbations, $1 \geq \beta \geq 0$, sets the degree of disorder in our system (Fig. 1 (a)) \citep{Walkama20,Haward21}. Flow in the ordered arrays ($\beta=0$) is periodic while disorder ($\beta>0$) leads to the emergence of high-velocity preferential flow pathways that surround the low-velocity pockets in the medium (Fig.~\ref{fig1} (b)). Geometric disorder broadens the velocity field distribution in the medium and leads to its deviation from the Gaussian distribution (Fig.~\ref{fig1} (c) and Figs. S1-3). 

We fill the medium with an aqueous solution of negatively charged 1 micron in size colloids with a background solute concentration $c_0$, and then flush the colloids out with a second aqueous solution with a solute concentration $c_1$. We denote the case with $c_1=c_0$ as our ``control" case, where solute gradients are absent. When $c_1>c_0$, the colloids will be “attracted” to the invading solute front, while for $c_1<c_0$, they will be “repelled” by the solute front; we therefore denote these two cases as “attractive” and “repulsive”, respectively. We then monitor the evolution of the colloidal density in a field of view downstream in the medium (Fig.~\ref{fig1} (a)). 

The transport of colloids in the absence of solute gradients, i.e. the control case, is modulated by the flow velocity distribution in the medium. In the ordered lattices, the colloid density decreases exponentially in time, $N/N_0\sim exp(-t/\tau)$, where $N$ represents the total number of colloids in the field of view with $N_0=N(t=0)$, and $\tau$ is the characteristic timescale of transport. This regime of transport is known as ``Fickian" (Supplementary Materials). In the presence of disorder, this Fickian regime is followed by a non-Fickian power-law regime with $N/N_0 \sim t^{-m}$ with $m<1$ \citep{Berkowitz06,Bijeljic11,Anna13,Puyguiraud21,Bordoloi22,Rajyaguru24}. This transition from the Fickian to non-Fickian behavior is due to the geometric disorder and its effect on the velocity distribution in the medium. In the low velocity and stagnant pockets of the medium, diffusion is the dominant escape route for the colloids, and therefore leads to their long residence time and the emergence of the power-law regime. 

Adding a “pinch of salt”, however, drastically changes this picture. In both ordered and disordered media, the attractive solute front removes the colloids much more effectively than the control case, shortening their macroscopic transport timescale through the medium, and even eliminating the non-Fickian regime of transport (Fig.~\ref{fig1} (d-g)). The repulsive solute front only leads to weak changes in the evolution of colloid density field compared to the control case. To understand the underlying mechanism of these observations, we need to probe the colloidal dynamics at the pore-scale. 

\begin{figure*}%[tbhp]
\centering
\includegraphics[width=1 \textwidth]{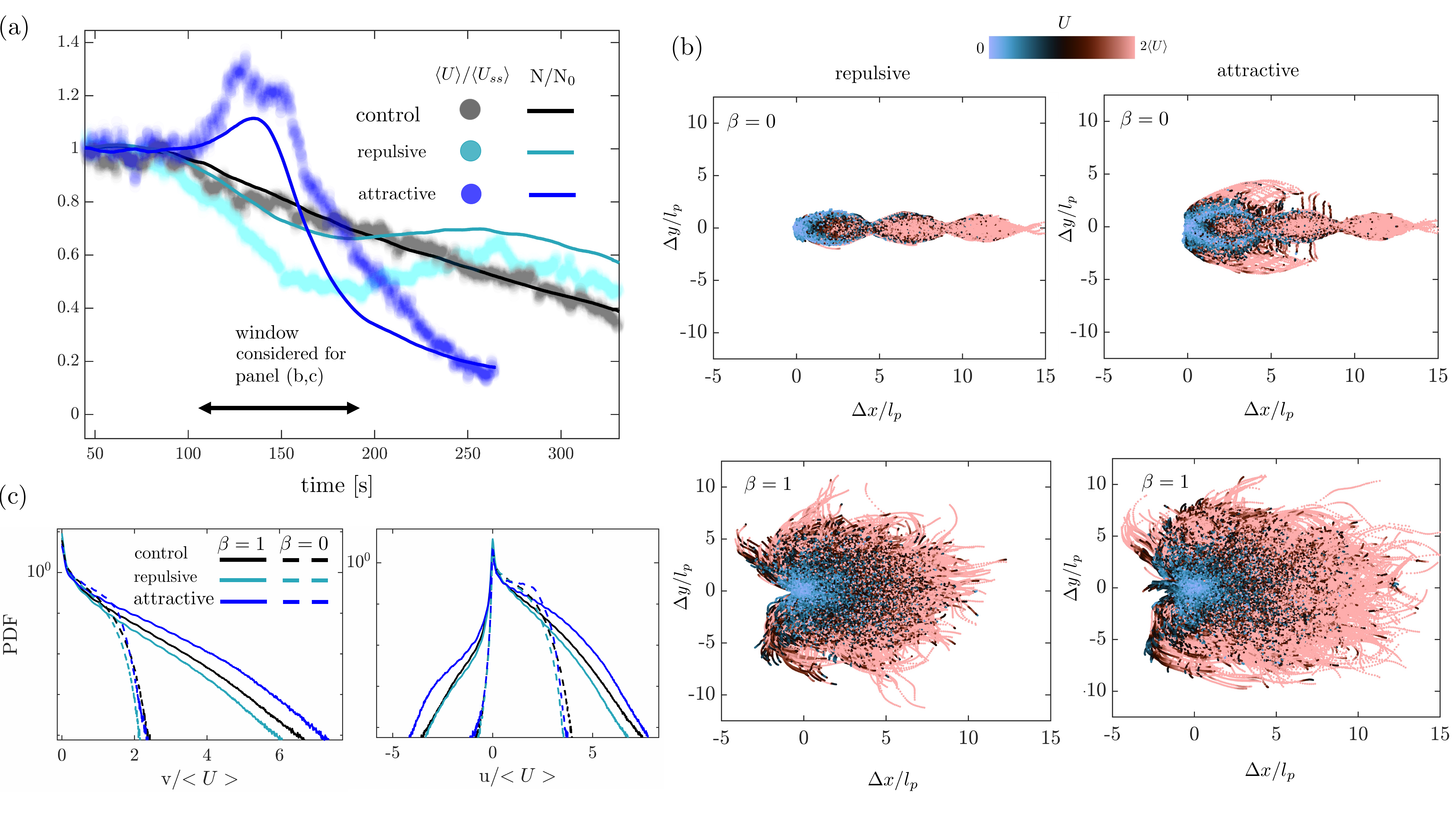}
\caption{Diffusiophoretic migration of colloids across the flow streamlines leaves a lasting fingerprint on the velocity distribution experienced by the colloids. (a) The temporal evolution of the normalized mean colloid velocity and density fields. In the control case, we observe a monotonic decrease in the colloid density field accompanied by a proportional decrease in the mean colloid velocity field. The attractive and repulsive cases, however, strongly deviate from this trend, showing a net transient increase in the mean colloid density and velocity in the attractive case and an opposite effect in the repulsive case. This panel corresponds to the disordered medium $\beta=1$. (b) Following the trajectories of the colloids, we observe a clear signature of the solute gradients, which suppress the transverse dispersion in the repulsive case while enhancing it in the attractive case. The displacement of colloids is normalized by the mean pore size $l_{p}$. (c) The attractive solute front pulls the colloids from the stagnant/low velocity zones toward the high-velocity streamlines, while the repulsive solute front has the opposite effect. These effects are evident in the probability density function of the colloidal velocity components, and is stronger in the disordered case due to the broader flow velocity distribution in the medium.}
\label{fig2}
\end{figure*}

The diffusiophoretic migration of colloids across the streamlines leads to an effective increase in the mean velocity of colloids in the attractive case and an effective decrease in the repulsive case (Fig.~\ref{fig2} (a)). In the control case, colloids follow their corresponding streamlines, which determine their travel time across the medium; their mean velocity therefore decreases proportionally to the colloid density field. This trend, however, is disrupted in the presence of solute gradients. In the attractive case, we observe a net increase in the mean velocity of the colloids, reflecting the arrival of fast colloids from upstream. This increase is followed by a sharp decrease in both the colloidal density and velocity fields as the fast upstream colloids traverse the medium and the slow colloids migrate toward the fast streamlines. On the other hand, in the repulsive case we observe a decrease in the mean velocity of the colloids, reflecting the phoretic migration from fast to slow streamlines, followed by the arrival of the slow upstream colloids, leading to a higher colloidal density compared to the control case (Fig.~\ref{fig2} (a)). These changes in the mean velocity of the colloids is an integral signature of the diffusiophoretic migration of colloids along the whole medium upstream of our field of view, pointing to the fact that minute changes in the colloidal trajectories at the pore-scale leave a lasting fingerprint on the macroscopic transport. 

Following the trajectories of colloids, we clearly observe a decrease in their transverse dispersion in the repulsive case and an opposite effect in the attractive case for both ordered ($\beta=0$) and disordered ($\beta=1$) media (Fig.~\ref{fig2} (b)). This effect is also evident in the velocity pdf of the colloids, which shows those in the attractive case experience a higher velocity in both the direction of flow and perpendicular to it due to their migration toward faster streamlines (Fig.~\ref{fig2} (c)). 

\begin{figure*}%[tbhp]
\centering
\includegraphics[width= 1 \textwidth]{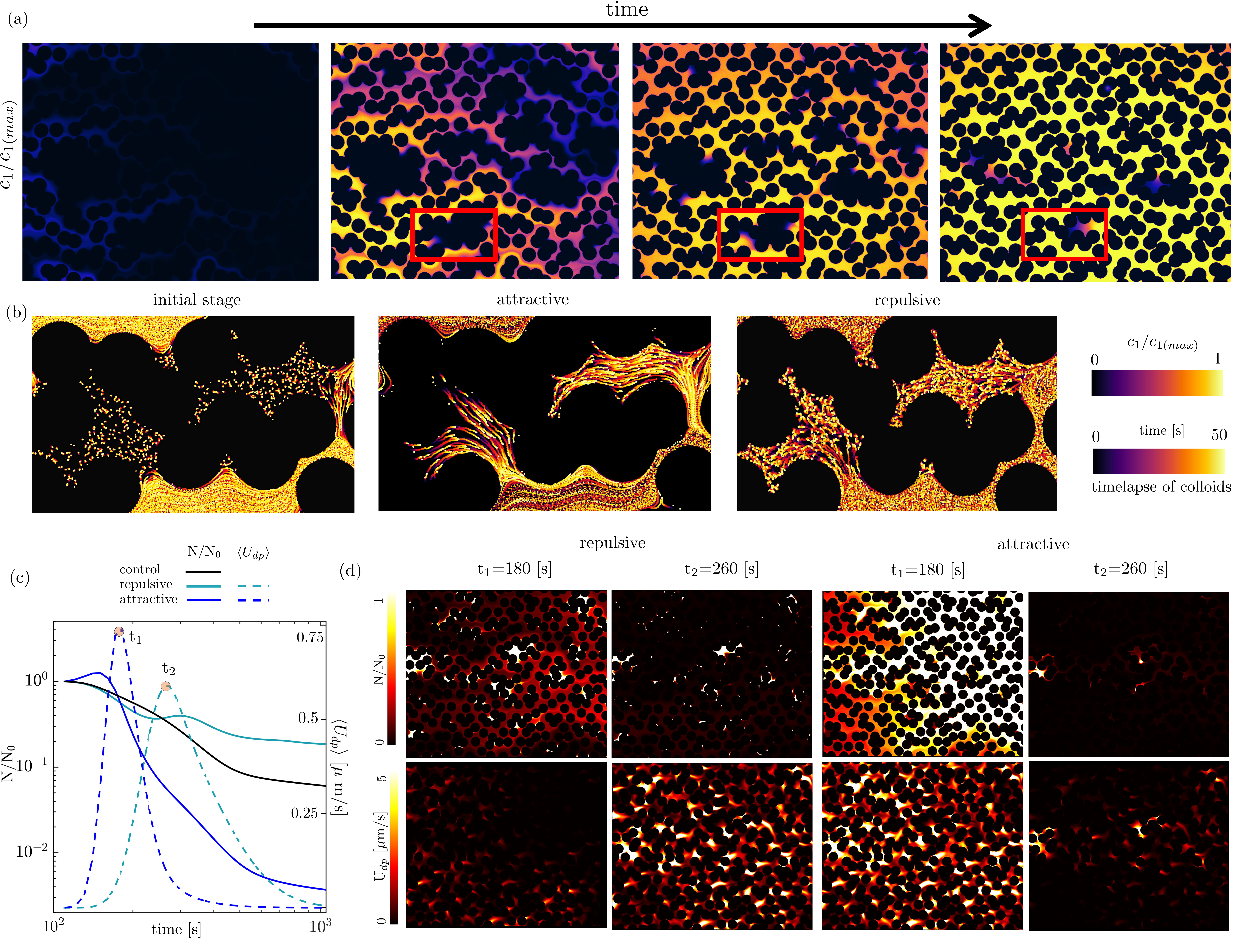}
\caption{Solute gradients drive the phoretic migration of colloids across the flow streamlines. (a,b) Using dual-channel imaging, we monitor the evolution of both the solute front tagged with a fluorescent probe, and the fluorescent colloids. The long-lasting solute gradients around the low-velocity pockets drive the phoretic migration of colloids out of these regions in the attractive case and into them in the repulsive case. (c) Using numerical simulations, we can disentangle the diffusiophoretic and advective velocity of the colloids, and monitor their concurrent evolution at both the pore and macroscopic scales. Surprisingly, the peak magnitude of the mean diffusiophoretic velocity in the medium is less than $\textrm{O}(1~\mu\textrm{m}/\textrm{s})$, which is two orders of magnitude weaker than the mean background flow velocity of $\textrm{O}(100~\mu\textrm{m}/\textrm{s})$. Yet, this ``weak" phoretic velocity leads to significant changes in the evolution of the colloid density field consistent with our experimental observations. (d) The evolution of the colloid density and phoretic velocity fields.}
\label{fig3}
\end{figure*}

To gain insight into how solute gradients at the pore-scale drive the diffusiophoretic migration of colloids, we use a dual-channel imaging technique to monitor both the solute front, which is tagged with a fluorescent dye, and the concurrent motion of the fluorescent colloids. Our experiments indeed demonstrate the correlation between the emergence of solute gradients at the pore-scale and the diffusiophoretic migration of colloids in/out of the stagnant fluid pockets (Fig.~\ref{fig3} (a,b)). However, we cannot experimentally disentangle the diffusiophoretic and advective components of the colloidal motion outside these stagnant pockets.  

We therefore resort to numerical simulations to probe the evolution of the diffusiophoretic velocity in the medium. The velocity field is governed by the Stokes equations, where we neglect the diffusioosmotic effects. The solute transport is governed by an advection-diffusion equation, and the colloid transport is coupled to the solute transport via the diffusiophoretic velocity:
\begin{align}
0&=-\nabla p + \mu \nabla^2 \mathbf{u},\nonumber\\
0&= \nabla \cdot \mathbf{u},\\
\frac{\partial c}{\partial t} & = D_s \nabla^2 c - \mathbf{u} \cdot \nabla c,\\
\frac{\partial n}{\partial t} & = D_p \nabla^2 n - \nabla \cdot \left[\left( \mathbf{u} + \mathbf{u}_{dp} \right) n\right], 
\end{align}
where $p$ is the fluid pressure, $\mathbf{u}$ is the flow velocity, and $c$ and $n$ are the solute and colloid density, respectively. The solute and colloid diffusion are denoted with $D_s$ and $D_p$, and the diffusiophoretic velocity is defined as $\mathbf{u}_{dp} = \Gamma_p \nabla \ln c$, where $\Gamma_p$ is the phoretic mobility of the colloids (Materials and Methods).

\begin{figure*}%[tbhp]
\centering
\includegraphics[width=1 \textwidth]{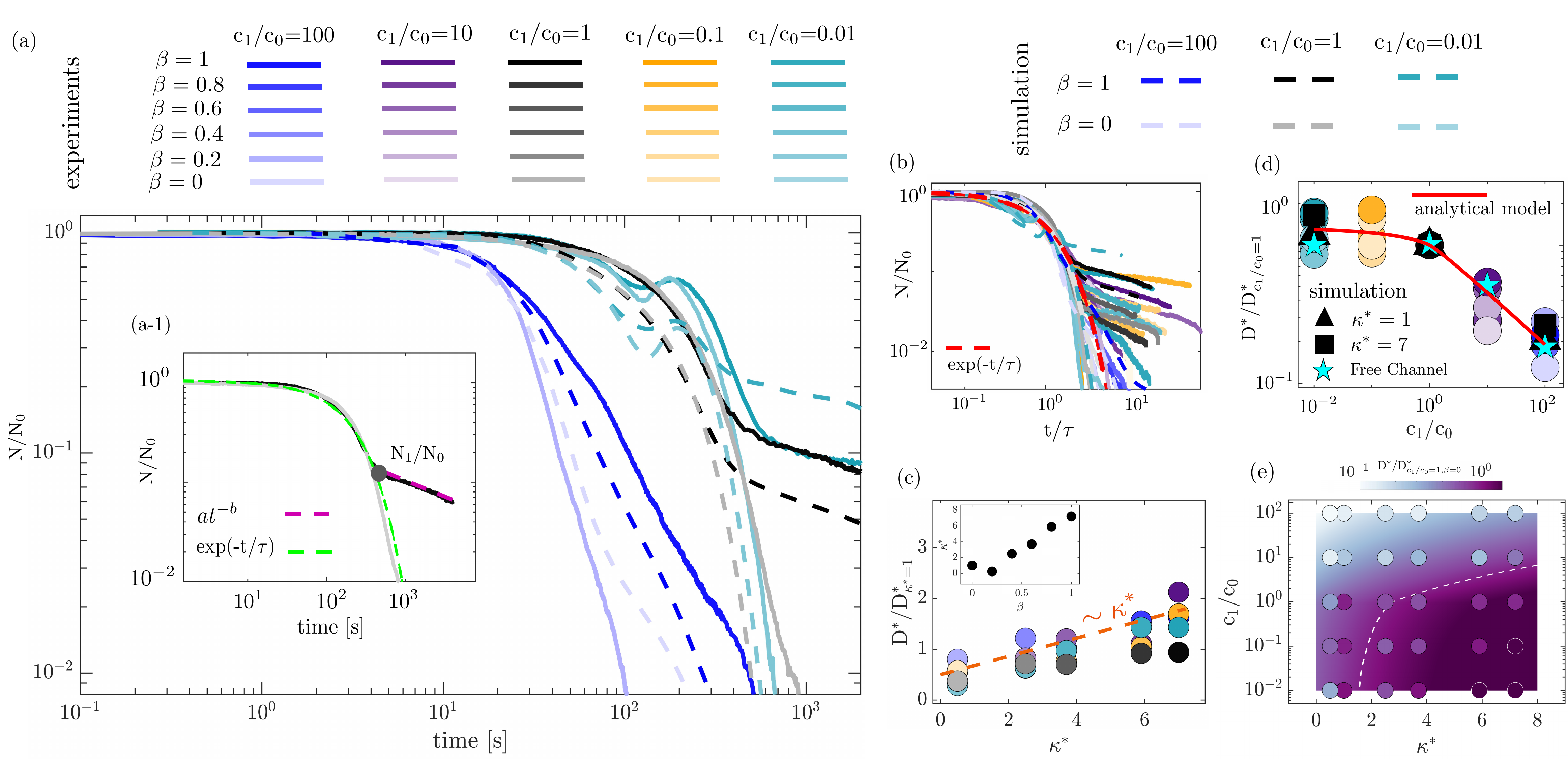}
\caption{Flow disorder and solute gradients shape the macroscopic dispersion of colloids. (a) The time evolution of the normalized colloid density field in the experiments (solid lines) and simulations (dashed lines) corresponding to the ordered $\beta=0$ and disordered $\beta=1$ media and attractive $c_1/c_0=100$, control $c_1=c_0$, and repulsive $c_1/c_0=0.01$ cases. The inset shows the exponential and power-law fits corresponding to the Fickian and non-Fickian regimes, respectively. (b) Normalizing the time with the characteristic exponential decay time $\tau$ of the Fickian regime, we obtain a reasonable collapse of the early-time dynamics of the data corresponding to 30 experiments and 6 simulations with different salt gradients $0.01 \leq c_{1}/c_{0} \leq 100$, and disorder strengths, $0\leq\beta\leq1$.  (c) The macroscopic dispersion coefficient defined as $D^* = U^2 \tau$ shows a linearly increasing trend with the normalized excess kurtosis of the transverse velocity distribution $\kappa^*=\kappa/\kappa_{\beta=0}$. The excess kurtosis characterized the flow disorder, and is a function of the geometric disorder parameter $\beta$ (inset). (d) The macroscopic dispersion coefficient in the attractive case decreases by an order of magnitude compared to the control case while it increases slightly in the repulsive case. We predict the influence of solute gradients using a simple model of non-diffusive colloids in a channel flow (red line). The stars represent the corresponding free-channel experiments with the same mean flow velocity and no obstacles. (e) The regime diagram of macroscopic dispersion, where circles represent the experimental data and the background colormap represents a semi-empirical model. The dashed line is a representative contour line.}
\label{fig4}
\end{figure*}

Our simulations show the same trend as our experiments, where the attractive solute front efficiently removes the colloids while the repulsive case remains close to the control case (Fig.~\ref{fig3} (c, d)). The arrival of the solute front leads to a transient peak in the diffusiophoretic velocity experienced by the colloids. Surprisingly, however, the peak magnitude of mean diffusiophoretic velocity in the medium is $\textrm{O}(1~\mu\textrm{m}/\textrm{s})$, which is two orders of magnitude weaker than the mean background flow velocity of $\textrm{O}(100~\mu\textrm{m}/\textrm{s})$. This observation confirms that the diffusiophoretic migration of colloids \textit{across the streamlines} is the key to the strong macroscopic changes in the evolution of colloid density field. 

Transport of colloids is therefore modulated by both flow disorder and solute gradients. We conduct experiments over a wide range of solute gradients and geometric disorders, monitoring the evolution of the colloidal density field (Fig.~\ref{fig4} (a)). Using the characteristic exponential decay timescale $\tau$ of the Fickian regime, we can obtain a reasonable collapse of the temporal evolution of the colloid density field at early times (Fig.~\ref{fig4} (b)). Using this timescale, we define an effective macroscopic dispersion for the colloidal field as $D^* = U^2 \tau$, where $U$ is the mean flow velocity (Supplementary Materials). 

We observe an increasing trend in the macroscopic dispersion with the flow disorder (Fig.~\ref{fig4} (c)). The macroscopic dispersion shows a non-monotonic behavior with respect to the geometric disorder parameter $\beta$ (Supplementary Materials), which is similar to the trend reported in a recent study \citep{Meigel22}. Our experiments show that this non-monotonicity correlates with the transverse velocity distribution as characterized by the excess kurtosis, which is a measure of the ``tailedness" of a distribution relative to the normal distribution, \text{$\kappa$} = $\frac{\mu_4}{\sigma^4} -\kappa_{\text{normal}}$, where $\mu_4$ and $\sigma$ are the fourth central moment and standard deviation of the distribution and $\kappa_{\text{normal}}$ is the kurtosis of normal distribution (Fig.~\ref{fig4} (c) inset). 

\begin{figure}%[tbhp]
\centering
\includegraphics[width=1 \textwidth]{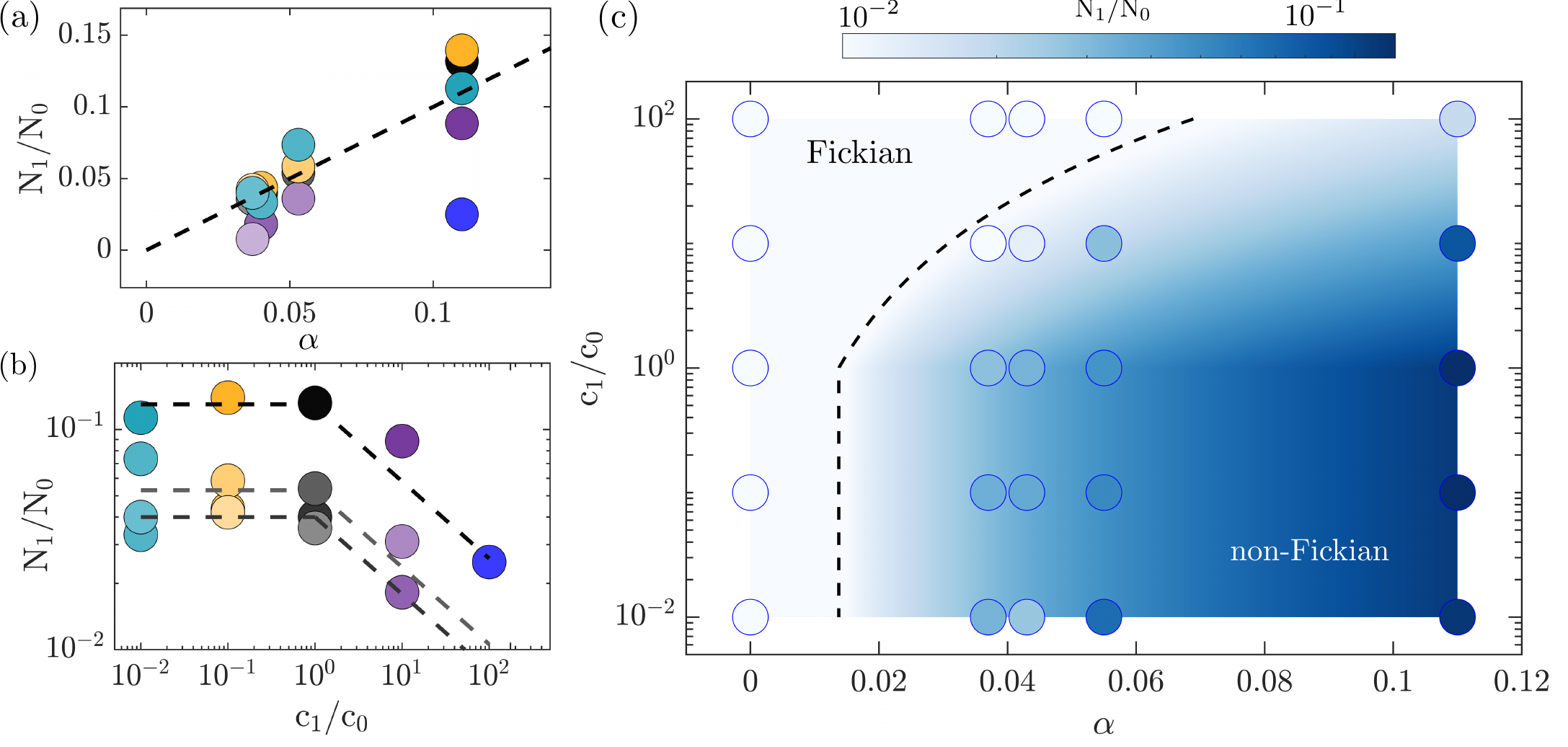}
\caption{Flow disorder and solute gradients modulate the transition between the Fickian and non-Fickian regimes of colloidal transport. (a) The residual density of colloids at the onset of non-Fickian regime is proportional to the area fraction of stagnant fluid pockets $N_1/N_0 \sim \alpha$. It is, however, also modulated by the solute gradients (b). While the repulsive front only weakly influences the residual colloid density, the attractive front removes the colloids from the stagnant pockets, weakening/eliminating the non-Fickian regime. Probing the transport in a dead-end pore, we find $N_1/N_0 \approx \alpha \left(c_1/c_0 \right)^{-\Gamma_p/D_s}$ (dashed line). (c) The regime diagram of macroscopic transport regime as characterized by the residual colloid density $N_1/N_0$, where circles represent the experimental data and the background colormap represents the model. The dashed line is a representative contour line.}
\label{fig5}
\end{figure}

The influence of flow disorder, however, is modulated by the solute gradients. The dispersion decreases in the attractive case by almost an order of magnitude compared to the control case (Fig.~\ref{fig4} (d)). This can be understood as an integral effect of the colloidal migration from the low to high velocity streamlines. On the other hand, the repulsive front leads to a weak increase in the macroscopic dispersion compared to the control case due to the push from high to low velocity streamlines. We note that this macroscopic dispersion is a measure of the longitudinal dispersion while the trajectories in Fig.~\ref{fig2} (b) highlighted the influence of solute gradients on the transverse dispersion of colloids. 

To gain insight into the influence of solute gradients on the macroscopic dispersion of colloids, we construct a model of the diffusiophoretic-driven dispersion of non-diffusive colloids in a channel flow. In the absence of diffusion, colloids can only phoretically move across the streamlines due to solute gradients. The velocity gradient between these streamlines in turn leads to the dispersion of colloids in the flow direction. Our model predicts that the attractive solute front significantly decreases the colloidal dispersion $D^*$ (Supplementary Materials), 
\begin{equation}
D^*/D^*_{\textrm{control}} \approx (c_1 /c_0 )^{-\Gamma_p/D_s}, 
\end{equation}
while the repulsive front slightly increases it, 
\begin{equation}
D^*/D^*_{\textrm{control}} \approx \left[ 7 \sqrt{3} - 3 (c_1 /c_0 )^{\Gamma_p/D_s}\right]/9,
\end{equation} 
which qualitatively captures the trend observed in our experiments (Fig.~\ref{fig4} (d)). 

The macroscopic dispersion of colloids through porous media is therefore modulated by both flow disorder and solute gradients as represented in the regime diagram of Fig.~\ref{fig4} (e). Here, the circles represent the experimental data, and the colormap represents the semi-empirical relationship $D^* \kappa^*$, where the empirical scaling with the flow disorder $\kappa^*$ is taken from Fig.~\ref{fig4} (c), and the dispersion $D*$ is predicted using the channel flow model above. 

The transition from the exponential Fickian regime to the power-law non-Fickian regime is determined by the transition in the dominant mechanism of transport of colloids from advection to diffusion. The diffusive mechanism is dominant in stagnant pockets, where advection is weak or absent. The area fraction of these pockets over the total available area of the medium is therefore expected to govern the extent of the non-Fickian regime. This is indeed the case in our experiments, which show the integral colloid density at the point of transition from Fickian to the non-Fickian regime is correlated with the area fraction of these stagnant pockets $\alpha$ (Fig.~\ref{fig5} (a)). This fraction, however, is modulated by the solute gradients. The attractive solute front can completely remove the trapped colloids, eliminating the non-Fickian regime, while the repulsive solute front leads to a weak increase in the colloid density in the stagnant pockets. 

To understand the influence of solute gradients on the fraction of trapped colloids, we probe the diffusiophoretic transport of non-diffusive particles in a dead-end pore. The attractive solute front draws the colloids out of this dead-end pore while the repulsive front pushes the colloids further into the pore. In the repulsive case, colloids traveling along the main flow channels can also be pushed into the dead-end pores and become trapped. However, this can only happen for colloids traveling ``slow enough" and ``close enough" to the pore. Considering the particle velocity in the main channel to be $U$, and its distance to the dead-end pore to be $\delta$, and the width of the pore to be $w$, for the trapping to occur we need $U_{dp} t_{adv} = U_{dp} (w/U) > \delta$, or $\delta/w < (U_{dp}/U) \approx 0.01$, which for a typical pore width of $w\approx 50~\mu\textrm{m}$, leads to $\delta <0.5~\mu\textrm{m}$. This simple scaling argument suggests that only colloids that are $\textrm{O}(1~\mu \textrm{m})$ away from the pore could become trapped due to the diffusiophoretic push toward the dead-end pores, and explains why the influence of repulsive front on the colloidal transport is much weaker than the attractive front \citep{Battat19}. 

For the attractive solute front, the residual particle density can be obtained as (Supplementary Materials)
\begin{equation}
N_1/N_0 \approx \alpha \left(c_1/c_0 \right)^{-\Gamma_p/D_s},
\label{eq:Fick}
\end{equation}
which matches our experimental observations (Fig.~\ref{fig5} (b)). 

The macroscopic regime of colloidal transport through porous media is therefore modulated by both flow disorder and solute gradients as represented in the regime diagram of Fig.~\ref{fig5} (c), where the circles represent the experimental data, and the colormap represents the Eq.~\eqref{eq:Fick} in the attractive regime and $N_1/N_0 \approx \alpha$ in the repulsive regime. 

Our observations highlight the strong influence of phoretic migration on determining the dispersion and transport of colloids in porous and crowded environments, demonstrating that solute gradients can strongly modulate the influence of geometric disorder on the transport, suppressing the influence of disorder on the macroscopic dispersion and eliminating the non-Fickian regime of transport that is commonly associated with the flow disorder. Surprisingly, we observe that even when the phoretic velocities are orders of magnitude weaker than the convective background flows, they can leave significant and long-lasting fingerprints on the macroscopic dispersion and transport of colloids due to the integral effect of the phoretic streamline crossing. The persistence of these trends in both ordered and disordered flow fields aligns with the reports on the signatures of diffusiophoretic migration on the macroscopic dispersion of colloids in cellular, chaotic, and even turbulent flows \citep{Volk14,Deseigne14,Schmidt16,Mauger16,Shukla17,Raynal18,Raynal19,Volk22}.

We therefore speculate that diffusiophoresis might contribute to a broad range of geophysical and biological flows. Beneath sea ice, salinity gradients emerge from the freezing and melting processes, generating unique microenvironments known as brine channels that harbor many organisms \citep{Arrigo14,Boetius15,Du24}. The dynamics of these gradients can influence the survival and distribution of marine organisms, including algae and bacteria, which form the base of the polar food web. In coastal ecosystems and estuaries, salinity gradients are of particular significance in controlling the transport and distribution of pollutants, such as microplastics \citep{Zhang17,Ouyang22,Wang22}. In subsurface flows, natural biogeochemical reactions, such as mineral dissolution and precipitation, can generate and sustain chemical and pH gradients \citep{Steefel05,Borch10,Li17,Rolle19}. These gradients not only might influence technological processes such as $\textrm{CO}_2$ mineralization \citep{Matter16} and hydrogen storage \citep{Heinemann21}, but also might play a role in maintaining the biodiversity around hydrothermal vents near the ocean floor \citep{Martin08,Ianeselli23}. Within the living cells, solute gradients could drive the large-scale transport of bio-macromolecules, contributing to the formation of membrane-less organelles and the regulation of biochemical pathways \citep{Sear19,Ramm21,Burkart22,Shandilya24,Doan24,Shim24,Hafner24}. The common ingredient in these phenomena is the presence of chemical gradients, crowding, and complex flow fields. The potential role and significance of diffusiophoresis in these problems remains to be uncovered.

\subsection*{Materials and Methods}
\subsection*{Experiments}
Microfluidic chips were fabricated using standard photolithography techniques. We designed hexagonal lattices with obstacle/post diameter of  $2 R = 165 \mu \textrm{m}$ and set the spacing between the posts, i.e., pore size to $d_p = 35 \mu \textrm{m}$, and introduced disorder by randomly perturbing the location of the obstacles with an amplitude $\beta L$, where $L=2R+d_p$ and $0\leq\beta\leq1$. The flow through the chip was fixed at 0.01 $\mu$l/s using a Harvard syringe pump. The images were recorded using a Nikon Ti-2 inverted microscope equipped with an ORCA-Fusion Digital CMOS camera with a spatial resolution of 2300 by 2300 pixels.  We used carboxylate-coated fluorescent polystyrene microspheres with diameter of 1 $\mu$m with excitation and emission wavelengths of 540 and 560 nm, respectively. We added the colloids to an aqueous solution of lithium chloride with molarity in the range of 0.01 to 100 mM. 

In a few experiments, we visualized the invading solute front by adding a dilute concentration (0.01 mM) of fluorescence sodium salt with excitation and emission wavelengths 460 and 515 nm, respectively. We note that the diffusivity of fluorescence is slightly different from the lithium chloride  $4.2 \times 10^{-10}$ m$^2$/s vs $1.37 \times 10^{-9}$ m$^2$/s, and therefore its evolution can only be considered an approximate proxy for the evolution of the lithium chloride front. Further, the presence of multiple ions could complicate the picture due to the coupled transport of ions \citep{Chiang14,Gupta19,Wilson20}. Therefore, we limited the use of fluorescence to a few representative experiments. We used dual-channel imaging with 10 frames/second to monitor the evolution of both colloids and the salt front in these experiments. All other experiments were done using a single channel imaging. 

To track the trajectories of the colloids, images were recorded at a rate of 64 frames per second, with a minimum particle displacement of 3 pixels per frame, covering a field of view measuring 2300x1800 pixels. We used an in-house code for the image processing in MATLAB together with the open-source software, TracTrac \citep{Heyman2019}.

\subsection*{Diffusiophoretic mobility}
The diffusiophoretic mobility of the colloids in a binary z-z electrolyte using the thin Debye layer approximation can be written as \citep{Prieve84,Anderson89}
\begin{equation}
    \Gamma_{p} = \frac{\epsilon}{\eta} \left( \frac{k_BT}{ze} \right)^2
    \left( \underbrace{\beta_s\frac{ze\zeta_{p}}{k_BT}}_{\text{electrophoresis}} + \underbrace{4\text{lncosh} \left( \frac{ze\zeta_{p}}{4k_BT} \right)}_{\text{chemiphoresis}} \right),
\end{equation}
, where $\eta$ is the viscosity of the fluid, $T$ is the temperature, $\epsilon$ is the permittivity of the medium, $k_B$ is the Boltzmann constant, $z$ is the electrolyte valence, $e$ is the elementary charge, $\zeta_p$ is the zeta potential of the colloid, and $\beta_s = \left(D_+ - D_-\right) / \left(D_+ + D_-\right)$ is the mobility difference of cation and anion, characterizing the electrophoretic strength. For LiCl, $D_+=1.03 \times 10^{-9} \textrm{m}^2/\textrm{s}$, $D_-=2.03 \times 10^{-9} \textrm{m}^2/\textrm{s}$. The colloids used in our experiments have a zeta potential $\zeta_p \approx -70 \textrm{mV}$ \citep{Shin16,Alessio22,Shim22b}, leading to the diffusiophoretic mobility of $\approx \Gamma_p=800 {\mu}m^2/s$. While zeta potential depends on the solute concentration, pH value, and counter-ion type \citep{Kirby04,Kirby04b}, we assume the zeta potential to be a constant in all the experimental scenarios for simplicity.

\subsection*{Numerical simulations}
The 3D numerical simulations were performed using the open-source software OpenFOAM, which employs Finite Volume (FV) schemes for the discretization of partial differential equations. The quasi-2D meshes were created with the mesh generator, snappyHexMesh and extrudeMesh. The steady flow field was first solved using simpleFoam solver. Then the transient transport processes of the solute and colloidal particles were computed with our customized solver, which considered the diffusiophoresis effect on particles. The initial solute concentration and particle density were set as 1 in the whole domain. At the inlet, the solute concentration was set to 0.01, 1 and 100, respectively for the repulsive, control and attractive cases, while the particle density was set to 0. Zero-gradient boundary conditions were employed at the outlet for both the scalars. In the gap direction, symmetry boundary condition was employed at the upper boundary. The other boundaries were treated as no-slip and no-flux walls.

\bibliography{main}% Produces the bibliography via BibTeX.

\end{document}